\documentclass[numreferences]{kluwer}
\usepackage{epsfig,graphicx}
\usepackage{amssymb}

\newcommand{\1}{{1\hspace{-3pt} \rm I}}

\newcommand{\AmS}{{\protect\the\textfont2
  A\kern-.1667em\lower.5ex\hbox{M}\kern-.125emS}}

\begin{document}
%\begin{article}

\begin{opening}
\title{Consequences of $t$-channel unitarity for the interaction of real and
virtual photons at high energies}

\author{E.~\surname{Martynov}\email{E.Martynov@guest.ulg.ac.be}\thanks{on leave from
Bogolyubov Institute for Theoretical Physics, Kiev.}}

\author{J.R~\surname{Cudell}} %\thanks{J.R.Cudell@ulg.ac.be}}

\author{G.~\surname{Soyez}}%\thanks{G.Soyez@ulg.ac.be}}

\institute{Universit\'e de Li\`ege, B\^at. B5-a, Sart Tilman,
B4000 Li\`ege, Belgium\\}

\runningtitle{Consequences of $t$-channel unitarity for the
interaction of real and virtual photons ...}

% If there are more authors at one institute, you should first
% use \author{...} for each author followed by \institute{...}.
\begin{abstract}
We analyze the consequences of $ t $-channel unitarity for
photon cross sections and show what assumptions are necessary to
allow for the existence of new singularities at $ Q^{2}=0 $ for
the $ \gamma p $ and $ \gamma \gamma $ total cross sections. For
virtual photons, such singularities can in general be present,
but we show that, apart from the perturbative singularity
associated with $ \gamma ^{*}\gamma ^{*}\rightarrow q\bar q $,
no new ingredient is needed to reproduce the data from LEP and
HERA, in the Regge region.
\end{abstract}

\end{opening}
\section{Introduction}

It is well known \cite{factorization} that due to unitarity one
can relate the amplitudes describing three hadronic elastic
processes $aa\rightarrow aa, ab\rightarrow ab, bb\rightarrow
bb$. Namely, if a simple Regge pole at $j=\alpha(t)$ contributes
in the $t$-channel for each of the above-mentioned processes,
the residues of the poles are factorized
$$
\beta_{aa\rightarrow aa}(t)\beta_{bb\rightarrow
bb}(t)=(\beta_{ab\rightarrow ab}(t))^{2}.
$$
However it is difficult to check directly such a relation.
Firstly, there are no experimental data for all three processes
(for example, $\pi \pi$ is missing, if one considers $\pi \pi,
\pi p$ and $pp$ scattering). Secondly, the best fit to the
hadronic cross section data is achieved in the models with
multiple Regge poles rather than with simple ones
\cite{COMPETE}. Factorization properties of multiple poles are
to be determined.

On the other hand, the DIS and total cross-section data
\cite{HERA} as well as the measurements of the $\gamma \gamma$
total cross section and of the off-shell photon structure
function $F_{2}^{\gamma}$ \cite{LEP} are available now. If
factorization is valid in the case of the photon amplitudes then
it can be checked for another set of related processes: $pp,
\gamma p, \gamma \gamma$ and, probably, for $pp, \gamma^{*}p,
\gamma^{*}\gamma^{*}$.

In this talk, we show how to derive the generalized
factorization for the partial amplitudes of the related
processes at an arbitrary, but common for these amplitudes,
$t$-channel Regge singularity. We also give arguments in favor
of its validity in the photon case and apply the new
factorization relations to describe $\gamma \gamma$ and
$\gamma^{*}\gamma^{*}$ cross sections.

\section{$t$-channel unitarity}

It is an old result that one can relate the amplitudes
describing three elastic processes $ aa\rightarrow aa $, $
ab\rightarrow ab $, $ bb\rightarrow bb $. The trick is to
continue these to the crossed channels $
a\overline{a}\rightarrow a\overline{a} $, $
a\overline{a}\rightarrow b\overline{b} $, $ b\overline{b}
\rightarrow b\overline{b} $, where they exhibit discontinuities
because of the $a$ and $b$ thresholds. One then obtains a
nonlinear system of equations, which can be solved. Working in
the complex $j$ plane above thresholds ($t>4m_a^2,\ 4m_b^2$),
and defining the matrix
\begin{equation}
T_{0}=\left( \begin{array}{cc}
A_{aa\rightarrow aa}(j,t) & A_{ba\rightarrow ba}(j,t)\\
A_{ab\rightarrow ab}(j,t) & A_{bb\rightarrow bb}(j,t)
\end{array}\right)
\end{equation}
one obtains\begin{equation} \label{solved} T_{0}={D\over \1-RD}
\end{equation}
with $ R_{km}=2i\sqrt{\frac{t-4m_{k}^{2}}{t}} \delta_{km}$ for
the case of two thresholds and $D=T_{0}^{\dagger }$. The latter
is made of the amplitudes on the other side of the cut. For any
$D$, equation (\ref{solved}) is enough to derive factorization:
the singularities of $T_0$ can only come from the zeroes of
\begin{equation}
\Delta=\det(1-RD).
\end{equation}
Taking the determinant of both sides of eq. (\ref{solved}), we obtain in the
vicinity of $\Delta=0$
\begin{equation}\label{factor1}
A_{aa\rightarrow aa}(j,t)A_{bb\rightarrow
bb}(j,t)-A_{ab\rightarrow ab}(j,t) A_{ba\rightarrow
ba}(j,t)={C\over\Delta},
\end{equation}
where $C$ is regular at
the zeroes of $\Delta$. As the l.h.s. is of order $1/\Delta^2$
we obtain the well-known factorization properties from eqs. (2)
and (4):
\\ $\bullet$ The elastic hadronic amplitudes have common singularities;
\\ $\bullet$ At each singularity in the complex $j$ plane, these amplitudes
factorise.

For isolated simple poles one obtains the usual well-known
factorization relations for the residues. However, it is
appropriate to mention here that the relation
\begin{equation}\label{factor2}
\lim _{j\rightarrow \alpha(t)}\left[A_{aa\rightarrow
aa}(j,t)-\frac{A_{ab\rightarrow ab}(j,t)A_{ba\rightarrow
ba}(j,t)}{A_{bb\rightarrow bb}(j,t)}\right ]=\mbox{finite
terms},
\end{equation}
where $\alpha(t)$ is the position of a zero of $\Delta$, is
valid not only for a simple pole but also for any common
$j$-singularity in the amplitudes. Moreover, it has a more
general form than just a relation between residues. To avoid a
misunderstanding we would like to note that ``any
$j$-singularity'' means a singularity of the full unitarized
amplitude rather than those partial singularities which are
produced \emph{e.g.} by $n$-pomeron exchange.

These equations are used to extract relations between the
residues of the singularities, which can be continued back to
the direct channel.

\subsection{Extension in the hadronic case}
We have extended \cite{cms} the above argument including all
possible thresholds, both elastic and inelastic. The net effect
is to keep the structure (\ref{solved}), but with a matrix $D$
that includes multi-particle thresholds. Furthermore, we have
shown that one does not need to continue the amplitudes from one
side of the cuts to the other, but that the existence of complex
conjugation for the amplitudes is enough to derive
(\ref{solved}) and consequently the factorization relations
(\ref{factor1},\ref{factor2}).

Hence there is no doubt that the factorization of amplitudes in the complex
$j$ plane is correct, even when continued to the direct channel.

If $ A_{pq}(j) $ has coinciding simple and double poles (at any
$t$ or {\it e.g.} colliding simple poles at $t=0$),
\begin{equation}
A_{pq}=\frac{S_{pq}}{j-z}+\frac{D_{pq}}{(j-z)^{2}},\end{equation}
one obtains the new relations
\begin{eqnarray}
D_{11}D_{22}&=&{\left( D_{12}\right) ^{2}},\nonumber\\
\label{double}
D_{11}^2S_{22}&=&{D_{12}(2S_{12}D_{11}-S_{11}D_{12})}.
\end{eqnarray}
In the case of triple poles
\begin{equation}
A_{pq}=\frac{S_{pq}}{j-z}+\frac{D_{pq}}{(j-z)^{2}}
+\frac{F_{pq}}{(j-z)^{3}},\end{equation}
the relations become
\begin{eqnarray}
F_{11}F_{22}&=&{\left( F_{12}\right) ^{2}},\nonumber\\
\label{triple}
F_{11}^2D_{22}&=&{F_{12}(2D_{12}F_{11}-D_{11}F_{12})},\\
F_{11}^3S_{22}&=&F_{11}F_{12}\left( 2S_{12}F_{11}-S_{11}F_{12}\right)
%\nonumber\\
+D_{12}F_{11}\left( D_{12}F_{11}-2D_{11}F_{12}\right)\nonumber\\
&+&D_{11}^2F_{12}^2.\nonumber
\end{eqnarray}

Although historically one has used $t$-channel unitarity to
derive factorization relations in the case of simple poles, it
is now clear \cite{DoLa} that a soft pomeron pole is not
sufficient to reproduce the $\gamma^*p$ data from HERA
\cite{HERA}. However, it is possible, using multiple poles, to
account both for the soft cross sections and for the DIS data
\cite{DM,CS}. We shall see later that relations (\ref{double},
\ref{triple}) enable us to account for the DIS photon-photon
data from LEP.

\subsection{The photon case}
For photons, due to the fact that an undetermined number of soft
photons can be emitted, two theoretical possibilities exist:\\
i) The photon cross sections are zero for any fixed number of
incoming or outgoing photons \cite{BN}. In this case, it is
impossible to define an S matrix, and one can only use unitarity
relations for the hadronic part of the photon wave function.
Because of this, photon states do not contribute to the
threshold singularities, and the system of equations does not
close. The net effect is that the singularity structure of the
photon amplitudes is less constrained. $\gamma p$ and
$\gamma\gamma$ amplitudes must have the same singularities as
the hadronic amplitudes, but extra singularities are possible:
in the $\gamma p$ case, these may be of perturbative origin, but
must have non perturbative residues. In the $\gamma\gamma$ case,
these singularities have their order doubled. It is also
possible for $\gamma\gamma$ to have purely perturbative
additional singularities.

\noindent ii) It may be possible to define collective states in
QED for which an S matrix would exist \cite{SM-QED}. In this
case, we obtain the same situation for on-shell photons as for
hadrons. However, in the case of DIS, virtual photons come only
as external states. Because they are virtual, they do not
contribute to the $t$-channel discontinuities, and hence the
singularity structure for off-shell photons is as described in
i).

\begin{figure}[htb]
\begin{center}
\includegraphics[scale=0.6]{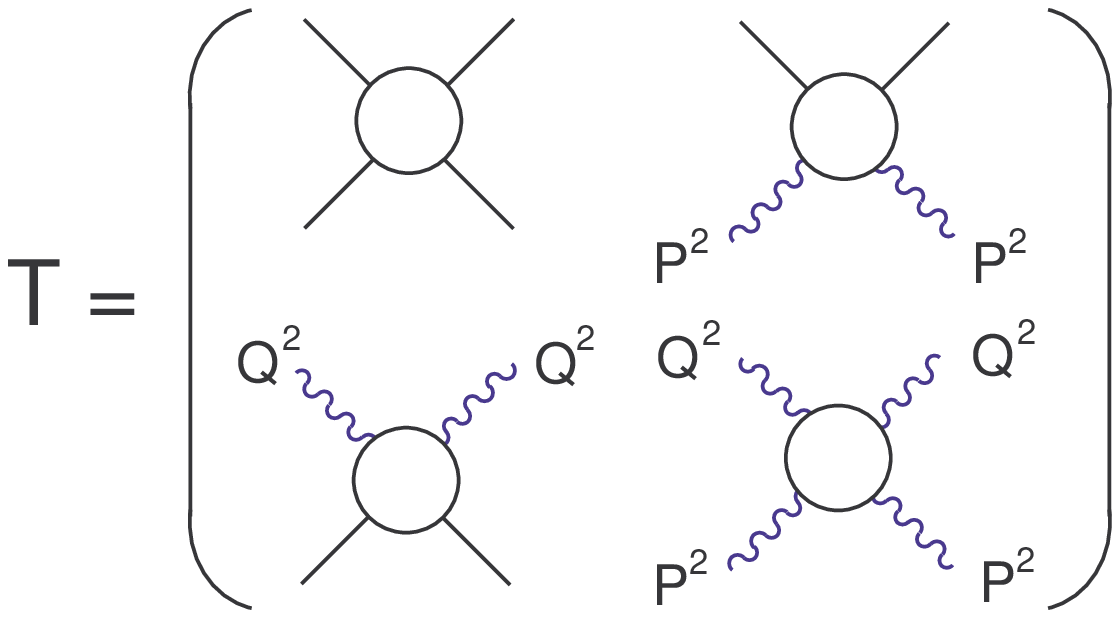}
\caption{Graphic representation of the matrix T in the case of
photons with virtualities $Q^2$ and $P^2$.}\label{fig:tmatrix}
\end{center}

\end{figure}

Let us consider the second possibility in more details and
define virtuality of photons as shown in Fig.\ref{fig:tmatrix}.

In the case of real photons ($Q^{2}=P^{2}=0$) we have obtained
for the matrix $T$ ($a\equiv p, b\equiv \gamma$) (see
\cite{cms}) the same expression as in the hadron case
(eq.(\ref{solved})). It means that there are no extra
singularities in the photon amplitudes besides those
contributing to $pp$ amplitude.

In the DIS case ($Q^{2}\neq 0, P^{2}=0$) we have
\begin{equation}\label{discase}
T(Q^{2},0)(\1-RD(0,0))=D(Q^{2},0),
\end{equation}
where $R$ is a diagonal matrix with elements
$R_{11}=2i\sqrt{(t-4m_{p})^{2}/t}$ and $R_{22}=2i$,
$D(Q^{2},P^{2})$ is expressed through
$T^{\dagger}(Q^{2},P^{2})$. Hence we see that all the on-shell
singularities must be present in the off-shell case, but we can
have new ones coming from the singularities of $ D(Q^{2},0) $.
These singularities can be of perturbative origin (\emph{e.g.}
the singularities generated by the DGLAP evolution) but their
coupling will depend on the threshold matrix $ R $, and hence
they must know about hadronic masses, or in other words they are
not directly accessible by perturbation theory.

In the case of $ \gamma ^{*}\gamma ^{*} $ scattering, we take
$Q^2\neq 0$ and $P^2\neq 0$, and obtain \cite{cms}
\begin{equation}
\label{Master2} T(Q^{2},P^{2})=D(Q^{2},P^{2})
+\frac{D(Q^{2},0)RD(0,P^{2})}{\1-RD(0,0)}.
\end{equation}
This shows that the DIS singularities will again be present,
either through $ \Delta =\mbox{det}(\1-RD(0,0)) $, or through
extra singularities present in DIS (in which case their order
will be different in $ \gamma \gamma  $ scattering, at least for
$Q^{2}=P^{2}$).

It is also possible to have extra singularities purely from $
D(Q^{2},P^{2}) $. \emph{A priori} these could be independent
from the threshold matrix, and hence be of purely perturbative
origin (\emph{e.g.} $ \gamma ^{*}\gamma ^{*}\rightarrow \bar{q}q
$ or the BFKL pomeron coupled to photons through a perturbative
impact factor).

We also want to point out that the intercepts of these new
singularities can depend on $ Q^{2} $, and as the off-shell
states do not enter unitarity equations, these singularities can
be fixed in $ t $. However, their residues must vanish as $
Q^{2}\rightarrow 0 $.

In the following, we shall explore the possibility that no new singularity
is present for on-shell photon amplitudes, and show that it is in fact
possible to reproduce present data using pomerons with double or triple poles
at $j=1$.

\section{Application to HERA and LEP}
For a given singularity structure, a fit to the $C=+1$ part of proton cross
sections, and to $\gamma^{(*)}p$ data enables one, via relations (\ref{solved}),
to predict the $\gamma^{(*)}
\gamma^{(*)}$ cross sections. Hence we have fitted \cite{cms} $pp$ and $\bar
p p$ cross sections and $\rho$ parameters, as well as DIS
data from HERA \cite{HERA}.
%\end{document}

The general form of the parametrizations which we used is given,
for total cross sections of $a$ on $b$, by the generic formula
$\sigma^{tot}_{ab}=(R_{ab}+H_{ab})$. The first term, from the
highest meson trajectories ($\rho$, $\omega$, $a$ and $f$), is
parametrized via Regge theory as
\begin{equation} R_{ab}= Y_{ab}^{+} \left({\tilde s}\right)^{\alpha _{+}-1} \pm
Y_{ab}^{-} \left({\tilde s}\right)^{\alpha _{-}-1}\label{lower}
\end{equation} with $\tilde s=2\nu/(1$ GeV$^2)$. Here the
residues $Y_+$ factorize. The second term, from the pomeron, is
parametrized either as a double pole \cite{DM,DM3}
\begin{eqnarray}
H_{ab}&=& D_{ab}(Q^2)\Re e\left[\log\left(1+\Lambda_{ab}(Q^2)\tilde
s\,^{\delta}\right)\right] \nonumber\\
&+&C_{ab}(Q^2)+(\tilde s\rightarrow -\tilde s)
\label{doubfit}
\end{eqnarray}
or as a triple pole \cite{CS}
\begin{equation}
H_{ab}=t_{ab}(Q^2)\left[ \log^2\left (
\frac{\widetilde{s}}{d_{ab}(Q^2)}\right ) + c_{ab}(Q^2) \right]
. \label{tripfit}
\end{equation}
It may be noted, in the double-pole case, that the parameter
$\delta$ is close to the hard pomeron intercept of \cite{DoLa}.
At high $Q^2$, because the form factor $\Lambda$ falls off, the
logarithm starts looking like a power of $\tilde s$, and somehow
mimics a simple pole. It may thus be thought of as a unitarized
version of the hard pomeron, which would in fact apply to hard
and soft scatterings.\\ In the triple-pole case, this is
accomplished by a different mechanism: the scale of the
logarithm is a rapidly falling function of $Q^2$, and hence the
$\log^2$ term becomes relatively more important at high $Q^2$.

\subsection{Results}
The details of the form factors entering (\ref{doubfit},
\ref{tripfit}) can be found in \cite{cms}. Such parametrizations
give $\chi^2/dof$ values less than 1.05 in the region
$\cos(\vartheta_t) \ge \frac{49}{2m_p^2},\ \sqrt{2\nu}\ge 7\
\mbox{GeV},\ x\leq 0.3,$\ $ Q^2\leq 150 \ {\rm GeV}^2$.
\begin{figure}[h]
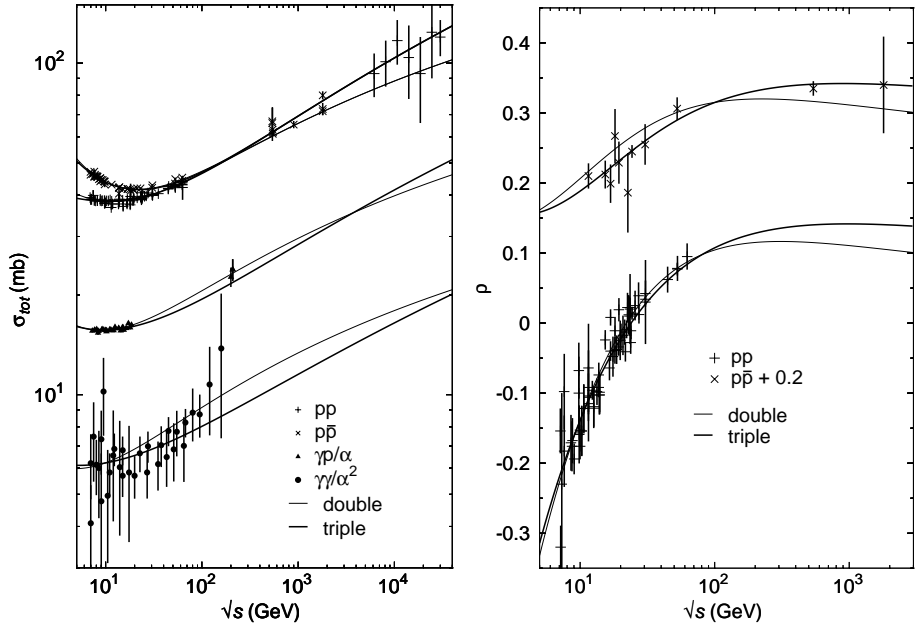

\centerline{
\includegraphics[width=6cm]{stot.eps}
\includegraphics*[width=6cm]{rho.eps}}
\caption{Fits to the total cross-sections and to the $\rho$
parameter. The thick and thin curves correspond respectively to
the triple-pole and to double pole cases.} \label{fig:strho}
\end{figure}
What is really new is that these forms can be extended to
photon-photon scattering, using relations (\ref{double},
\ref{triple}). The total $\gamma\gamma$ cross section is well
reproduced (see Fig.~\ref{fig:strho}) and the de-convolution
using PHOJET is preferred.

The fit to $F_2$ has quite a good $\chi^2$ as well. We have
checked that one can easily extend it to $Q^2\approx 400$
GeV$^2$ for the triple pole, and to $Q^2\approx 800$ GeV$^2$ in
the double-pole case. It is interesting that one cannot go as
high as in ref. \cite{CS}. This can be attributed either to too
simple a choice for the form factors, or more probably to the
onset of perturbative evolution.

\begin{figure}
\includegraphics[scale=0.75]{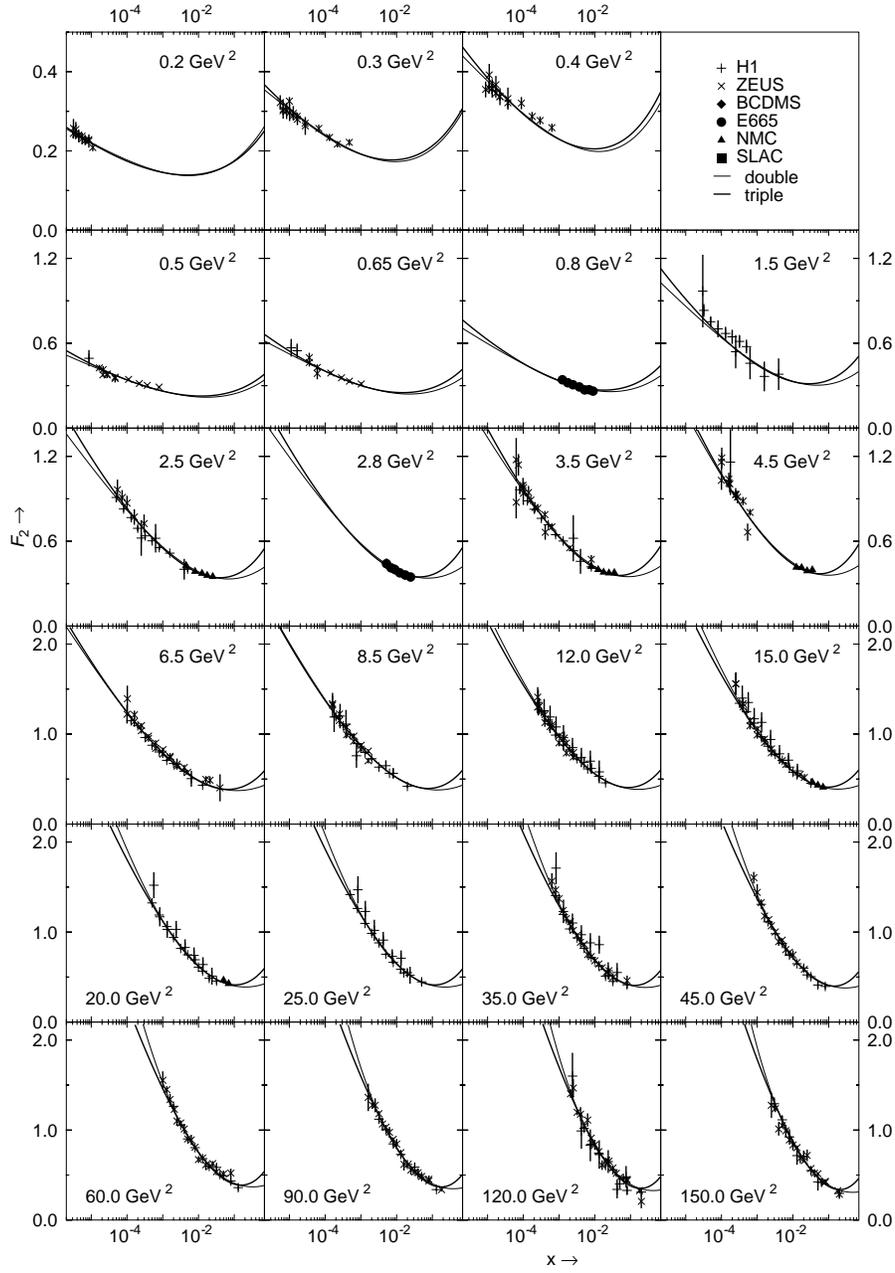}
\vskip 0.5cm
\caption{Fits to $F_2^p$. We show only graphs for which there
are more than 6 experimental points, as well as the lowest $Q^2$
ones. The curves are as in Fig.~\ref{fig:strho}}
\label{fig:f2q2}
\end{figure}

Fig.~\ref{fig:f2q2} shows the $F_2^p$ fit for some selected
$Q^2$ bins (figures for other  $Q^{2}$ bins can be found in
\cite{cms}). As pointed out before, our fits do reproduce the
low-$Q^2$ region quite well, but predict total cross sections on
the lower side of the error bands. Hence the extrapolation to
$Q^2=0$ of DIS data does not require a hard pomeron.

\begin{figure}[h]
\caption{Fits to $F_2^\gamma$. The thick and thin curves
correspond respectively to the triple-pole and to the
double-pole cases. The data are from \cite{LEP}.}
\label{fig:f2gam}
\centerline{\includegraphics*[width=12cm]{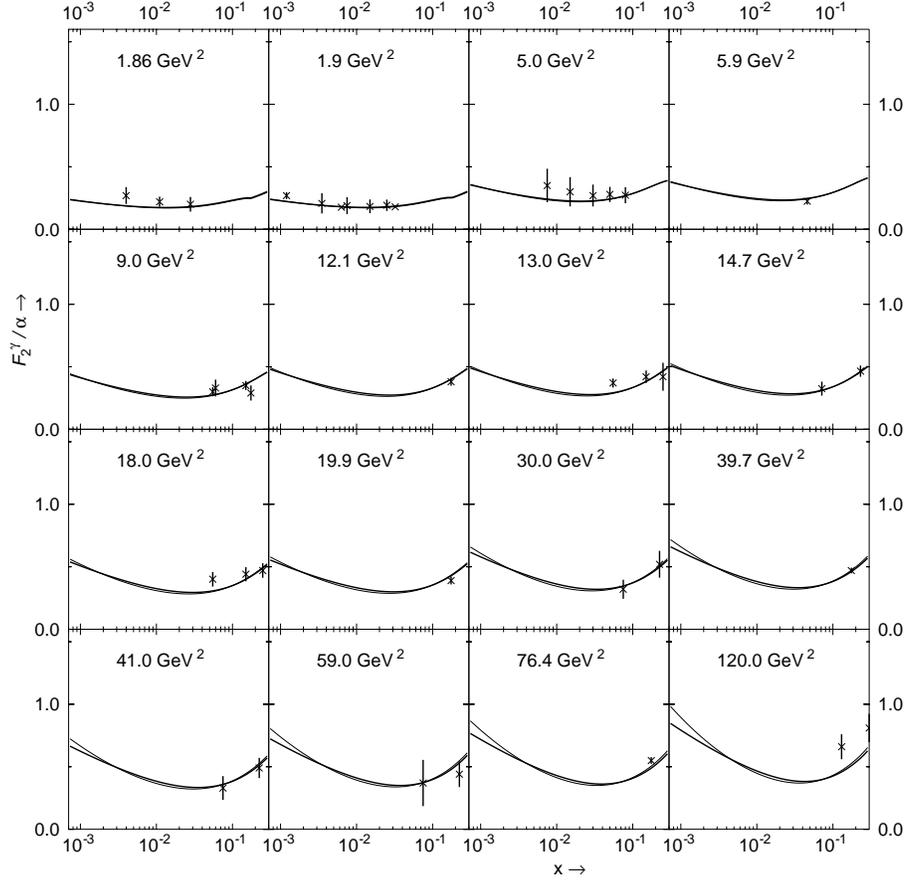}}
\end{figure}

For photon structure functions, one needs to add one singularity
at $j=0$ corresponding to the box diagram \cite{Budnev}, but
otherwise the $\gamma\gamma$ amplitude is fully specified by the
factorization relations. One can see from Fig.~\ref{fig:f2gam}
that the data on photon structure are well reproduced by both
parametrizations.
%(see Fig.~4).
%Fig.~\ref{gammaDIS} shows that DIS data
%are well reproduced by both parametrizations.

\begin{figure}[h]
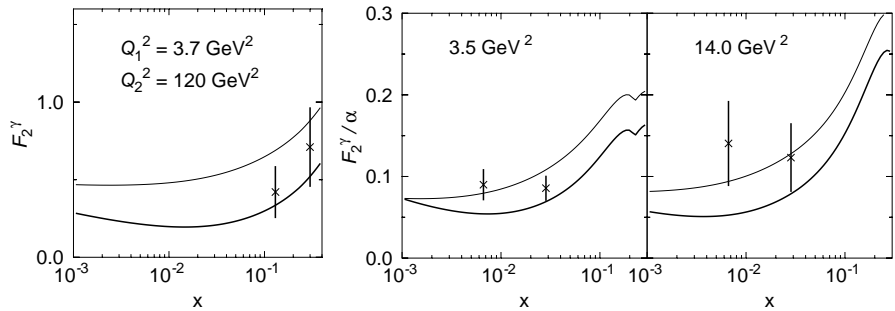

\includegraphics*[width=4.3cm]{g37120.eps}
\includegraphics*[width=7.5cm]{gsame.eps}
\caption{Fits to $F_2^\gamma$ for  nonzero asymmetric values of
$P^2$ and $Q^2$ and for $P^2=Q^2$. The curves are as in
Fig.~\ref{fig:f2gam}. The data are from \cite{LEP}.}
\label{fig:gsame}
\end{figure}

Even more surprisingly, it is possible to reproduce the $\gamma^*
\gamma^*$ cross sections when both photons are off-shell, as shown in
Fig.~\ref{fig:gsame}. This is the place where BFKL singularities may
manifest themselves, but as can be seen such singularities are not needed.

In conclusion, we have shown that $t$-channel unitarity can be
used to map the regions where new singularities can occur, be
they of perturbative or non-perturbative origin. Indeed, we have
seen that although hadronic singularities must be universal,
this is not the case for $F_2^p$ and $F_2^\gamma$, as DIS
involves off-shell particles. Nevertheless, up to $Q^2=150$
GeV$^2$, the data do not call for the existence of new
singularities, except perhaps the box diagram. In the case of
total cross sections, this suggests that it is indeed possible
to define an $S$ matrix for QED.

For off-shell photons, our fits are rather surprising as the
standard claim is that the perturbative evolution sets in quite
early. This evolution is indeed allowed by $t$-channel unitarity
constraints: it is possible to have extra singularities in
off-shell photon cross sections, which are built on top of the
non-perturbative singularities. But it seems that Regge
parametrisations can be extended quite high in $Q^2$ without the
need for these new singularities.

Thus it is possible to reproduce soft data ({\it e.g.} total
cross sections) and hard data ({\it e.g.} $F_2$ at large $Q^2$)
using a common $j$-plane singularity structure, provided the
latter is more complicated than simple poles. Furthermore, we
have shown that it is then possible to predict $\gamma\gamma$
data using $t$-channel unitarity. How to reconcile such a simple
description with DGLAP evolution, or BFKL results, remains a
challenge.

%\begin{figure}[h]
%\centerline{
%\includegraphics*[width=5.cm]{lowq2.eps}%}
%\caption{Fits to $F_2^p$ at low $Q^{2}$. We show only graphs for
%which there are more than 6 experimental points, as well as the
%lowest $Q^2$ ones. The curves are as in Fig.~\ref{fig:strho}}
%\label{fig:f2lowq2}
%\end{figure}
%\begin{figure}[h]
%\centerline{
%\includegraphics*[width=5.cm]{highq2.eps}%}
%\caption{Fits to $F_2^p$ at low $Q^{2}$. Notations are the same
%as in Fig.~\ref{fig:f2lowq2}} \label{fig:f2highq2}
%\end{figure}}
%\end{rotate}

%\end{article}

\begin{thebibliography}{9}
\bibitem{factorization}V.N.~Gribov and I.~Ya.~Pomeranchuk: 1962, {\it Phys. Rev.
Lett.} \textbf{8}, p. 4343; J. Charap and E.J. Squires: 1962,
{\it Phys. Rev.} {\bf 127}, p. 1387 .
\bibitem{COMPETE} COMPETE collaboration, J.~R.~Cudell
{\it et al.}: 2002, {\it Phys.\ Rev.\ } {\bf D65}, p. 074024
[arXiv:hep-ph/0107219];
%%CITATION = HEP-PH 0107219;%%
2002, {\it Phys. Rev. Lett.} {\bf 89}, p. 201801
[arXiv:hep-ph/0206172].
%%CITATION = HEP-PH 0206172;%%
\bibitem{HERA} H1 collaboration:
%T. Ahmed et al. , Nucl. Phys. B {\bf 439}, 471
%(1995); %%CITATION = NUPHA,439,471;%%
%S. Aid et al. , Nucl. Phys. B {\bf 470}, 3 (1996); %%CITATION = NUPHA,470,3;%%
C.~Adloff
%Nucl. Phys. B {\bf 497}, 3 (1997); %%CITATION = NUPHA,497,3;%%
%Eur. Phys. J. C {\bf 13}, 609 (2000); %%CITATION = EPHJA,13,609;%%
{\it et al.}: 2001 {\it Eur. Phys. J. } {\bf C21}, p. 33;  %%CITATION = EPHJA,21,33;%%
ZEUS collaboration:
%M. Derrick et al. , Zeit. Phys. C {\bf 72}, 399 (1996);
%%CITATION = ZEPYA,72,399;%%
%J. Breitweg et al. , Phys. Lett. B {\bf 407}, 432 (1997);
%%CITATION = PHLTA,407,432;%%
%Eur. Phys. J. C {\bf 7}, 609 (1999); %%CITATION = EPHJA,7,609;%%
%Nucl. Phys. B {\bf 487}, 53 (2000); %%CITATION = NUPHA,487,53;%%
S.~Chekanov {\it et al.}: 2001, {\it Eur. Phys. J.}  {\bf C21},
p. 443,
%%CITATION = EPHJA,21,443;%%
and references therein.
\bibitem{LEP}L3 Collaboration:
M.~Acciarri {\it et al.}:
%Phys.\ Lett.\ B {\bf
%408} (1997) 450; %%CITATION = PHLTA,B408,450;%%
% Phys.\ Lett.\ B {\bf 436} (1998) 403; %%CITATION = PHLTA,B436,403;%%
% Phys.\ Lett.\ B {\bf 447} (1999) 147; %%CITATION = PHLTA,B447,147;%%
% Phys.\ Lett.\ B {\bf 453} (1999) 333; %%CITATION = PHLTA,B453,333;%%
% Phys.\ Lett.\ B {\bf 483} (2000) 373 [arXiv:hep-ex/0004005];
%%CITATION = HEP-EX 0004005;%%
2001, {\it Phys.\ Lett.\ }  {\bf B519}, p. 33;
%%CITATION = HEP-EX 0102025;%%
OPAL Collaboration:
%R.~Akers {\it et al.},
%Z.\ Phys.\ C {\bf 61} (1994) 199; %%CITATION = ZEPYA,C61,199;%%
%K.~Ackerstaff {\it et al.}
%Phys.\ Lett.\ B {\bf 411} (1997) 387 [arXiv:hep-ex/9708019];
%%CITATION = HEP-EX 9708019;%%
%Phys.\ Lett.\ B {\bf 412} (1997) 225 [arXiv:hep-ex/9708028];
%%CITATION = HEP-EX 9708028;%%
% Z.\ Phys.\ C {\bf 74} (1997) 33; %%CITATION = ZEPYA,C74,33;%%
G.~Abbiendi {\it et al.}:
%Eur.\
%Phys.\ J.\ C {\bf 14} (2000) 199 [arXiv:hep-ex/9906039];
%%CITATION = HEP-EX 9906039;%%
% Eur.\ Phys.\ J.\ C {\bf 18} (2000) 15 [arXiv:hep-ex/0007018];
%%CITATION = HEP-EX 0007018;%%
%Eur.\ Phys.\ J.\ C {\bf 24} (2002) 17 [arXiv:hep-ex/0110006];
%%CITATION = HEP-EX 0110006;%%
2002, {\it  Phys.\ Lett.\ } {\bf B533}, p. 207, and references
therein.
%%CITATION = HEP-EX 0202035;%%
\bibitem{cms}J.R.~Cudell, E.~Martynov and G.~Soyez,
arXiv:hep-ph/0207196.
%%CITATION = HEP-PH 0207196;%%
\bibitem{DoLa}
A.~Donnachie and P.~V.~Landshoff: 1998, {\it Phys.\ Lett.\ }
{\bf B437}, p. 408;
%%CITATION = HEP-PH 9806344;%%
{\it Phys.\ Lett.\ } {\bf B533}, 277 (2002).
%%CITATION = HEP-PH 0111427;%%
\bibitem{DM}P.~Desgrolard and E.~Martynov: 2001, {\it Eur.\ Phys.\ J.\ }
{\bf C22}, p. 479 [arXiv:hep-ph/0105277].
%%CITATION = HEP-PH 0105277;%%
\bibitem{CS}J.~R.~Cudell and G.~Soyez: 2001, {\it Phys.\ Lett.\ } {\bf B516}, p. 77
[arXiv:hep-ph/0106307].
%%CITATION = HEP-PH 0106307;%%
%\bibitem{total}
%The data on the total $\gamma p$ cross-section prior to HERA are extracted
%from http://pdg.lbl.gov. The HERA data come from:
%%CITATION = NONE;%%
%H1 collaboration, S. Aid et al. , Zeit. Phys. C {\bf 69}, 27
%(1995);
%%CITATION = ZEPYA,69,27;%%
%ZEUS Collaboration; S. Chekanov et al. , Nucl. Phys. B {\bf 627},
% (2002).
%%CITATION = NUPHA,627,3;%%
\bibitem{BN}F.~Bloch and A.~Nordsick: 1937, {\it Phys. Rev. } {\bf 52}, p. 54.
\bibitem{SM-QED}P.~P.~Kulish and L.~D.~Faddeev: 1970, {\it Theor.\
Math.\ Phys.\ } {\bf 4}, p. 745 [1970, {\it Teor.\ Mat.\ Fiz.\ }
{\bf 4}, p. 153];
%%CITATION = TMPHA,4,745;%%;
\\D.~Zwanziger: 1976, {\it Phys.\ Rev.\ } {\bf D14}, p. 2570;
%%CITATION = PHRVA,D14,2570;%%;
\\E.~Bagan, M.~Lavelle and D.~McMullan: 2000, {\it Annals Phys.\ } {\bf 282}, p. 503
[arXiv:hep-ph/9909262];
%%CITATION = HEP-PH 9909262;%%
R.~Horan, M.~Lavelle and D.~McMullan: 2000, {\it J.\ Math.\
Phys.\ } {\bf 41}, p. 4437 [arXiv:hep-th/9909044], and
references therein.
%%CITATION = HEP-TH 9909044;%%
\bibitem{DM3}P.~Desgrolard, A.~Lengyel and E.~Martynov: 2002,
{\it JHEP } {\bf 0202}, 029 [arXiv:hep-ph/0110149].
%%CITATION = HEP-PH 0110149;%%
\bibitem{Budnev}V.~M.~Budnev, I.~F.~Ginzburg, G.~V.~Meledin and
V.~G.~Serbo: 1974, {\it Phys.\ Rep.\ } {\bf 15}, p. 181.
%%CITATION = PRPLC,15,181;%%
\end{thebibliography}
\end{document}